\documentclass[12pt]{iopart}

\usepackage{graphicx}
\usepackage{epsfig}
\usepackage{iopams}
\newcommand{\sgn}{{\rm sgn}\,}

\begin{document}

\title{Survival Probability of a Doorway State in regular and chaotic environments}

\author{Heiner Kohler$^{(1)}$, Hans--J{\"u}rgen Sommers$^{(1)}$ and  
Sven \AA berg$^{(2)}$}

\address{
$^{(1)}$Fakult{\"a}t f{\"u}r Physik, Universit\"at Duisburg--Essen, Duisburg, 
Germany \\
$^{(2)}$Matematisk Fysik, Lunds Universitet, Lund, Sweden}
\ead{heinerich.kohler@uni-due.de}

\begin{abstract}
We calculate survival probability of a special state which couples randomly to a regular or chaotic environment. The environment is modelled by a suitably chosen random matrix ensemble. The exact results exhibit non--perturbative features as revival of probability and non--ergodicity. The role of background complexity and of coupling complexity is discussed as well.   
\end{abstract}

\maketitle

\section{Introduction}
\label{sec1}

The stability of a special prepared quantum state weakly coupled to a continuum
is a subject of considerable interest in quantum information theory \cite{nie00}, nuclear
physics, mesoscopic, quantum chaos (see \cite{gor06,jaq08} and references therein). Giant resonances are
collective excitations of the nucleons, which are approximate but not exact
eigenstates of the complicated many--body Hamitonian. They are an example of a
special state coupled weakly to a continuum picked from nuclear physics.
Constructed superscar states in chaotic quantum billiards, considered recently
\cite{bog06} fall into the same class. The special state might also be implemented
mesoscopically, for instance by an electronic state on a small conducting island
(quantum dot), which is weakly coupled to one (or in a non--equilibrium
situation to two) continua of the leads \cite{akk06}. 
Given a particle in a quantum state $|\psi(0)\rangle$ at time $t=0$ the
likelyhood to find the particle after some time $t$ in the same state is
measured by the survival probability 
\begin{equation}
P(t) \ =\ |\langle \psi(t)|\psi(0)\rangle|^2. 
\end{equation}
The simplest approximation of $P(t)$ is found perturbatively by F(ermi's) G(olden)
R(ule) $P(t) \ =\ e^{-\Gamma t}$, where $\Gamma$ is the inverse decay time, related to the
mean coupling strength of the special state to the background. 
Corrections to this simple exponential decay become important, when $\Gamma$ 
has the same order of magnitude as the mean level spacing of the
background Hamiltonian.

In Ref.~\cite{pri94} corrections to the FGR--law, were calculated in a 
mesoscopic system, where the special state is sitting on a quantum dot which is  weakly coupled to a
reservoir. Due to these corrections the decay of the electronic state in the
dot is never complete. Instead the system preserves a memory of the original
state. The corrections to FGR, which are in nature very similar to the weak localisation correction to classical transport,
give rise to non--ergodic behavior.
In the same spirit, in Ref.~\cite{gut09}  weak localisation corrections to the FGR
behavior were calculated using new semiclassical techniques, which became available recently \cite{sie01,mue04}. 

Corrections to the FGR--law can be addressed in a more generic setting within a random matrix
model. In this model a special state $|\psi(0)\rangle$ $=$ $|s\rangle$ couples to a
large reservoir of states via randomly chosen coupling parameters $V_\mu$,
$\mu=1,\ldots N$. The coupling parameters are chosen either real or complex. 
The dynamics of the reservoir is either chaotic, modelled by a
random matrix chosen from a Gaussian RMT ensemble or regular, with
Poissonian eigenvalue statistics. This RMT model has been addressed in
\cite{gru97,flo98} and corrections to FGR regime were found. 

Expanding the special state in eigenstates of the Hamiltonian survival
probability can be written as
\begin{equation}
P(t) \ =\ \sum_{n,m} |\langle s|m\rangle|^2 |\langle s|n\rangle|^2
e^{i(E_n-E_m)t} \ . \label{surprobdef}
\end{equation} 
For uncorrelated eigenvalues and expansion coefficients $\langle s|m\rangle$ the
ensemble average (denoted by a bar) can be performed for both sums separately. As
a result 
\begin{equation}
\label{FGRapp}
\overline{P(t)}= |\overline{p(t)}|^2 \ , 
\end{equation}
where $p(t)$ is the Fourier transform of the {\em local
density of states} LDOS 
\begin{equation}
\rho(E) \ =\ \sum_n |\langle s|n\rangle|^2 \delta(E-E_n+E_s) \ 
\end{equation}
around the energy $E_s$ of the special state. As was pointed out already by Weisskopf and Wigner \cite{wei30} the smooth part
of $\rho(E)$ is under very general assumptions of Lorentzian shape
\begin{equation}
\label{BWshape}
\overline{\rho(E)} \ =\ \frac{1}{\pi}\frac{\Gamma/2}{(E-E_s)^2+(\Gamma/2)^2} \ ,
\end{equation}
from which FGR is recovered. We call the approximation implied in Eq.~(\ref{FGRapp}) {\em D(rude) B(oltzmann)--approximation}, due to its formal similarity to the approximations made in the derivation of the Drude Boltzmann law of conductivity \cite{akk06}.

Assuming a constant mean level spacing $D$ in an energy region around the special state, from Eq.~(\ref{surprobdef}) one obtains $P(\infty)$ $=$ ${\rm IPR}$, where 
\begin{equation}
{\rm IPR} \ =\ \sum_m|\langle s|m\rangle|^4 \ =\ D \int dE \rho^2(E)
\end{equation}
is the inverse participation ratio of the special state in the basis of the eigenstates of the full Hamiltonian. In the DB--approximation $\overline{\rho^2}$ $\simeq$ $\overline{\rho}^2$  and the mean IPR can be estimated by $ \overline{\rm IPR} \simeq D/(\pi \Gamma)$, which shows that survival probability will not decay to zero, if $\Gamma $ and $D$ are of 
the same order of magnitude.   

Following Ref.~\cite{gru97} one can obtain corrections to the DB approximation due to energy correlations
 as follows: Writing the expansion coefficients $|\langle s|n\rangle|^2$ $=$ $\overline{\rho}(E_n) +$ $\delta\rho(E_n) $ as 
the sum of a smooth function of $E_n$ and of a fluctuating part, the averaged survival probability can be written as the sum of two contributions as well
\begin{equation}
\overline{P(t)} \ =\ \int dE dE^\prime \overline{\rho(E)}\, \overline{\rho(E^\prime)} e^{i(E-E^\prime)t} R_2((E-E^\prime)/D)
               + \overline{\delta P(t)}\ , 
\end{equation} 
where $R_2$ $=$ $\sum_{n,m} \overline{\delta(E-E_n)\delta(E^\prime-E_m)}$ is the averaged energy--energy correlator of the background Hamiltonian. The approximation made in Ref.~\cite{gru97} consists in neglecting the fluctuating part $\overline{\delta P(t)}$.  Using standard results of random matrix theory \cite{meh04} $R_2$ is the sum of a $\delta$--like, a connected and of an unconnected contribution. Likewise, survival probability is written as a sum of three contributions   
\begin{equation}
\label{gru}
\overline{P(t)} \ \simeq \ e^{-\Gamma t} +  \frac{D}{\pi\Gamma}  - 
                   \int\limits_{-\infty}^{\infty} d t^\prime e^{-\Gamma|t-\frac{2\pi}{D}t^\prime|} b_2(t^\prime) 
\end{equation} 
where $b_2$ is the two--level form factor. The last term in Eq.~(\ref{gru}) accounts for the energy correlations of the bath. 
The result (\ref{gru}) gives an intuitive insight how energy correlations of the background Hamiltonian give rise to corrections of the FGR--law and as we will see for strong coupling $\Gamma \gg D$ and for a chaotic background it predicts qualitatively the correct behavior. 

However, it is easy to see that for small $\Gamma$ or in the case where correlations are absent Eq.~(\ref{gru}) is not correct even qualitatively (for instance the saturation value  $\overline{P(\infty)}$ $=$ $ D/(\pi \Gamma)$ exceeds one for $\Gamma<D/\pi$).

Thus the question, whether Eq.~(\ref{gru}) describes sufficiently the weak localisation corrections to the DB--approximation has to be answered negatively. In the present work we therefore calculate $\overline{P(t)}$ for the random matrix model mentioned above exactly for a chaotic as well as for a regular background. We will see that the exact result differs qualitatively from the predictions of Eq.~(\ref{gru}). For instance, we find that a revival of survival probability, predicted by Eq.~(\ref{gru}) only for fairly strong couplings and for a chaotic background, occurs for weak coupling and for a regular background as well. More general, the energy statistics of the background turn out to have little influence on the survival probability. Instead, we find that the nature of the coupling coefficient is crucial. As a rule of thumb, for constant mean coupling strength survival probability is always lower for complex coupling coefficients than for real ones. 

We provide exact analytic expressions for the average IPR, which interpolate the power law decay for strong couplings to the small coupling regime. 

On the technical level, we use powerful results for averages over characteristic polynomials, which have become available recently \cite{bor04}. This allows us to circumvent a long and complicated supersymmetric calculation. This elegant shortcut is possible for complex couplings of the doorway state to the background, where we derive exact analytic results. For real coupling coefficients we resort to numerics. 

The paper is organized as follows:
In Sec. \ref{sec21} and in Sec. \ref{sec22} we define the random matrix model and fix the notation. 
In Sec. \ref{mldos} we outline how the Lorentz shape of the LDOS comes about in the present random matrix model. The calculation of survival probability for regular background as well as for a GUE and for a GOE background is presented in Sec.~\ref{sec3}.
Finally the results are discussed and summarized in Sec.~\ref{dis}.

\section{Definition of the Doorway model}
\label{sec21}

The model to be discussed here stems from nuclear physics~\cite{boh69} and
is also often used in other fields~\cite{guh98}. For the convenience of
the reader and to define our notation, we compile its salient features. 

The total Hamiltonian $H$ consists of three parts, the Hamiltonian
$H_s$ for the doorway states,
the Hamiltonian $H_b$ describing the $N$ background states, where
$N$ will eventually be taken to infinity, and the interaction $V$ coupling 
the two classes of states. Often, there is only one relevant doorway state or the spacing between the
doorway states is much larger than their spreading widths. In the present work we focus on this situation, leaving 
the interesting case of many doorway states to future work. 
Hence, we have
\begin{eqnarray}
H&=& H_s+ H_b +V\nonumber\\
H&=& H_0 +V\nonumber\\
&=& E_{s}|s\rangle\langle s| + \sum_{\nu=1}^N E_{\nu}|b_\nu\rangle\langle b_\nu| + \sum_{\nu=1}^N \Big( V_{\nu} |s\rangle\langle b_\nu| + {\rm h.c.}\Big) \ .
\label{e1}
\end{eqnarray}
For the matrix elements of the interaction, we make the assumptions 
$\langle b_\nu|V|b_\mu\rangle=0$ and
$\langle b_\nu|V|s\rangle=V_{\nu}$ for any $\mu$, $\nu$.

Resembling the situation in most systems, we put the  doorway state 
$|s\rangle$ in the center of the background spectrum. It interacts with the $N$ 
surrounding states. Without loss of generality, we may set $E_s=0$.
The eigenequations for the uncoupled Hamiltonian $H_0$ are then
\begin{equation}
H_s|s\rangle = 0
\qquad {\rm and} \qquad
H_b|b_\nu\rangle = E_\nu|b_\nu\rangle \ .
\label{e2}
\end{equation}
 
We assume that the interaction matrix elements
are Gaussian distributed random variables. We distinguish the two cases of complex ($\beta=2$) or real ($\beta=1$) matrix elements $V_\nu$. Introducing the $N$--component vector
$V$, the corresponding distribution is
\begin{equation}
P_i(V) \ = \  \left(\frac{\beta}{2\pi v^2}\right)^{\beta N/2}
                \exp\left(-\frac{\beta}{2v^2}V^\dagger V\right) \ . 
\label{e9a}
\end{equation}
As discussed in the introduction, we are interested in the situation where the mean coupling strength is of the same order of magnitude as the mean level spacing. We define the dimensionless parameter 
\begin{equation}
\lambda \ = \ \frac{\sqrt{\langle V^\dagger V}\rangle}{\sqrt{N}D} \ =\ \frac{v}{D} \ ,
\label{e9b}
\end{equation}
where $D$ is the mean level spacing of the background
states in the center of the band~\cite{boh69,guh98}. The distribution $P_i(V)$ 
is chosen such that $\lambda$ is independent of $\beta$. 

We distinguish two cases for the background dynamics. 
Regular dynamics of the background Hamiltonian $H_b$ is modelled by uncorrelated eigenvalues 
\begin{equation}
P_b(H_b) \ = \ \prod_{\nu=1}^N p_b(E_\nu) \ .
\label{e13}
\end{equation}
Chaotic dynamics is modelled by Gaussian random matrix ensembles given by the distribution function
\begin{equation}
P_b(H_b) \ = \left(\frac{\beta_b}{2\pi}\right)^{\frac{N}{2}} \left(\frac{\beta_b}{\pi}\right)^{\frac{\beta_b N (N-1)}{4}} \exp\left(-\frac{\beta_b}{2}\tr H_b^2\right) \ ,
\label{e20}
\end{equation}
where $H_b$ is either real symmetric ($\beta_b=1$) modelling time reversal invariant background dynamics, or Hermitean ($\beta_b=2$), modelling background dynamics with broken time reversal invariance. The probability distribution (\ref{e20}) yields a mean level spacing $D=\sqrt{\pi^2/2N}$ in the band center, which is independent of $\beta_{b}$.
We denote the average over both the interaction matrix elements and the background Hamiltonian by a bar
\begin{equation}
\overline{(\ldots)} \ =\ \int d[H_b] P_b(H_b) \int d[V] P_i(V)(\ldots) \ .
\end{equation}

\section{Fidelity and Survival probability of the Doorway state}
\label{sec22}

We define the echo operator
\begin{equation}
\label{echo}
M_\lambda \ =\ e^{iH t}e^{-iH_0 t} \ .
\end{equation}
Fidelity amplitude $f_\lambda(t)$ is defined as the expectation value of the echo operator with respect to a given initial state.
Here we are interested in the Doorway state $|s\rangle$ as initial state, i.e. an eigenstate of the unperturbed system.
Since $E_s|s\rangle$ $=0$ the average fidelity amplitude can be written as
\begin{equation}
\overline{f_\lambda(t)} \ =\ \overline{\left< s| M_\lambda(t)|s\right>} \ = \ \overline{\left< s| e^{-iH t}|s\right>} .
\end{equation}
As mentioned in the introduction, fidelity amplitude is then the Fourier transform of the local density of states $\rho(E)$.
Likewise, fidelity (often called {\em Loschmidt echo}) $F_\lambda(t)$, defined as the modulus square of the fidelity amplitude, becomes identical with the survival probability $P(t)$
\begin{eqnarray}
\overline{F_\lambda(t)} & = & \overline{\left< s|e^{-iH t} |s\right>\left< s|e^{iH t}|s\right>} 
                         \ \equiv\ \overline{P(t)}\ .
\end{eqnarray}
In the following we mainly stick with the notion of survival probability, keeping in mind that in the present situation fidelity and survival probability are synonyms.

When we expand fidelity in eigenstates of the full Hamiltonian according to Eq.~(\ref{surprobdef}), we see that in the limit of infinite large times the fidelity approaches the inverse participation ratio IPR of the special state in the basis of the eigenstates of the system. We therefore also define the mean inverse participation ratio 
 \begin{eqnarray}
 \label{F31}
\overline{{\rm IPR}_\lambda} \ &\equiv& \ \overline{\sum_m |\langle s|m\rangle|^4} \ =\ \overline{F_\lambda(\infty)} \ ,
\end{eqnarray}
where the sum goes over exact eigenstates of the full Hamiltonian $H$.
The task is to calculate $\overline{f_\lambda(t)}$ and $\overline{F_\lambda(t)}$ exactly in the large $N$ limit for various choices of the background Hamiltonian and in particular the corrections to the DB--approximation.

\section{Calculation of the mean local density of states}
\label{mldos}
We first prove that the average local density of states $\overline{\rho(E)}$ in the present case indeed has the Lorentz shape. We write
\begin{eqnarray}
\rho(E) & =& \sum_m|\langle s|m\rangle|^2\delta(E-E_m)\nonumber\\
                   & =& \frac{1}{\pi} {\rm Im} \left< s\left| \frac{1}{H-E-i\epsilon}\right|s\right>\nonumber\\
                   &=& \frac{1}{\pi} {\rm Im} \frac{\det(H_0-E)}{\det(H-E-i\epsilon)}\ , 
\end{eqnarray}
where we used Kramer's rule in the step from the second to the third equation. For a chaotic background the average can be taken most conveniently by a mapping onto a supersymmetric matrix model. Using standard steps \cite{stoe99} one arrives at
\begin{eqnarray}
\label{maint}
\overline{\rho(E)} & =&\frac{1}{\pi} {\rm Im}\int d[\sigma]\exp\left(-\frac{\beta_b}{2}{\rm Str}(\sigma+E)^2\right)
                        {\rm Sdet}^{-\frac{\beta_b N}{2}}\left(\sigma\right) \nonumber\\
                   &&\quad\qquad \overline{{\rm det}^{-\beta_b/2}\left(\frac{V^\dagger V}{\sigma_{\rm BB}} +E\right)}\ ,
\end{eqnarray}   
where the bar now denotes an average over the coupling coefficients only. In Eq.~(\ref{maint}) $\sigma$ is a $2\times 2$ (GUE, $\beta_b=2$) respectively $4\times 4$ (GOE, $\beta_b=1$) supermatrix of the form
\begin{eqnarray}
\left(\begin{array}{cc}
a_1& \lambda_1^*\cr
\lambda_1& ia_2	
\end{array}\right)&,\quad & {\rm GUE}\nonumber\\ 
\left(\begin{array}{cccc}
a_1&a_2& \lambda_1^*&-\lambda_1\cr
a_2&a_3& \lambda_2^*&-\lambda_2\cr	
\lambda_1&\lambda_2&ia_4&0\cr
\lambda_1^*&\lambda_2^*&0&ia_4
\end{array}\right)&,\quad& {\rm GOE} \ .
\end{eqnarray}
The matrix entries in Latin letters denote real commuting integration variables. The matrix entries in Greek letters denote complex anticommuting integration variables. The infinitesimal volume elements $d[\sigma]$ are products of the differentials of all independent integration variables. The integration domain of the real commuting variables is the real axis. The so--called Boson--Boson block $\sigma_{\rm BB}$ is the upper left block of commuting variables. The matrix integral Eq.~(\ref{maint}) can be solved in one step with a saddlepoint approximation. For energies close to the center of the band the saddle points are $\sigma \simeq \pm i \sqrt{N/2}$ and thus
\begin{eqnarray}
\label{maint1}
\overline{\rho(E)} & =& \frac{1}{\pi}{\rm Im}\overline{\left(E-i \frac{2 D}{\pi} V^\dagger V\right)^{-1}} \ .
\end{eqnarray}
This result shows that LDOS has the form of a $\delta$--spike unless the perturbation is classically small, i.~e. of the order of the mean level
spacing. The Gaussian average over the coupling coefficients finally yields the Lorentz distribution
\begin{equation}
\overline{\rho(E)} \ =\ \frac{1}{\pi }\frac{\Gamma/2}{E^2+(\Gamma/2)^2}\ 
\end{equation}
with the crucial relation 
\begin{equation}
\Gamma = 2\pi\lambda^2 D
\end{equation}
between spreading width, mean perturbation strength and mean level spacing.
For a regular environment the LDOS was calculated  for instance in \cite{mel95} yielding the same result.

\section{Calculation of mean Fidelity/Survival Probability}
We now turn to the main task: the calculation of survival probability of the doorway state. In order to calculate the mean fidelity/survival probability, we write $\overline{F_\lambda(t)}$ as  
\begin{eqnarray}
\label{F00}
F_\lambda(t) &=&\frac{1}{\pi^2} \int dE_1 \int dE_2 \exp\left(i(E_1-E_2)t\right)
                            \rho(E_1+i\epsilon)\rho(E_2-i\epsilon)\nonumber\\ 
              &=& \frac{1}{\pi^2} \int dE_1 \int dE_2 \exp\left(i(E_1-E_2)t\right)\nonumber\\ 
               &&    \frac{\det(H_0-E_1)}{\det(H-E_1-i\epsilon)} 
                     \frac{\det(H_0-E_2)}{\det(H-E_2+i\epsilon)}\ ,
\end{eqnarray}
where we used again Kramer's rule. Evaluation of the determinant in the denominator yields                            
\begin{eqnarray}
F_\lambda(t) &=& \int dE_1 \int dE_2 \exp\left(i(E_1-E_2)t\right) \delta\left(E_1+\sum_\mu\frac{|V_\mu|^2}{E_\mu-E_1}\right)\nonumber\\
              &&\qquad    \delta\left(E_2+\sum_\mu\frac{|V_\mu|^2}{E_\mu-E_2}\right) \ .
\end{eqnarray}
We observe that $F$ is normalized by $F_\lambda(0)=1$ (see \cite{gru97,koh09}). This allows us to extract at this point of the calculation the constant term $F_\lambda(0)$ from the integral and to average over $F_\lambda(t)-1$ 
instead of $F(t)$ directly.     
After a Fourier transformation of the delta distributions we find for the average 
 \begin{eqnarray}
\overline{F_\lambda(t)} &=&1+ \int dE_1 \int dE_2 \int \frac{dk_1}{2\pi}\int 
                              \frac{dk_2}{2\pi}\left[\exp\left(it(E_1-E_2)\right)-1\right]\nonumber\\
&&\exp\left(-i\frac{k_1-k_2}{2}(E_1-E_2)-i\frac{k_1+k_2}{2}(E_1+E_2)\right)\nonumber\\
&& \overline{\exp\left(-i\sum_\mu\frac{|V_\mu|^2 k_1}{E_\mu-E_1} - i\sum_\mu\frac{|V_\mu|^2k_2}{E_\mu-E_2}\right)} \ .
\end{eqnarray}
Since on the unfolded scale the mean level spacing of the background is constant, we can assume that the unfolded average does not depend on $E_1+E_2$. This allows us to perform the integral over the mean energy $(E_1+E_2)/2$ and, trivially, over $(k_1+k_2)$.  We find 
 \begin{eqnarray}
 \label{F0}
\overline{F_\lambda(\tau)} &=&1+ \int ds \int \frac{dk}{2\pi} 
                              \left[\exp\left(i\tau s)\right)-1\right]\exp\left(-i k s\right) R(k,s)\ ,
\end{eqnarray}
where 
\begin{eqnarray}
\label{rks}
R(k,s) &=& \overline{\exp\left(-i\sum_\mu\frac{|V_\mu|^2 ks}{E^2_\mu-(Ds/2)^2}\right)}\ .
\end{eqnarray}
We introduced  the dimensionless time $\tau= D t$, measured in units of Heisenberg time $\tau_{\rm H}= D^{-1}$.
It is useful to take the average over the Gaussian distributed coupling coefficients at this stage of the calculation
\begin{eqnarray}
R(k,s) &=& \int \prod_{\mu=1}^N\left\{\frac{d[V_\mu]^\beta}{(2\pi v^2/\beta)^{\beta/2}}\right.\nonumber\\ 
     &&\left.\overline{
        \exp\left[-\frac{\beta}{2v^2}
        \left(1+\frac{2 i k s D^2\lambda^2}{\beta(E_\mu^2-(Ds/2)^2)}\right)|V_\mu|^2\right]}\right\}\nonumber\\
       &=&\overline{\left(\frac{\det(H_b^2-(Ds/2)^2)}
             {\det (H_b^2-(Ds/2)^2+  i k s D^2\lambda^2)}\right)^{\beta/2}} \ , \label{ch10} 
\end{eqnarray}
where the bar denotes the average over the background Hamiltonian $H_b$ only, which has still to be performed. In the expression (\ref{ch10}) it becomes evident why the case of real coupling coefficients $V_\mu \in {\mathbb R}$, $\beta=1$  is analytically more difficult than the case of complex coupling $\beta=2$. For $\beta=2$ the expression  (\ref{ch10}) is an average over a rational ratio of products of characteristic polynomials. Much information has been gathered about these averages in the last decades \cite{bre00,bre00a,bre01,str03,bai03,fyo03,bor04,dei00}, whereas little is known about averages over irrational functions of characteristic polynomials as encountered in the case $\beta=1$. 

In the subsequent analysis we therefore restrict ourselves to complex coupling and set $\beta=2$ from now on. In the case of real coupling we recur to numerics. We distinguish the cases of a regular background $R_{\rm Poisson}(k,s)$, and GOE or GUE distributed chaotic background $R_{\rm GOE}(k,s)$, $R_{\rm GUE}(k,s)$. 
We are able to calculate all three averages exactly in the large $N$ limit.

\subsection{Survival Probability for a regular Background}
\label{sec23} 
We first consider a regular, Poisson distributed, background. 
For Poisson distributed eigenvalues the average in Eq.~(\ref{ch10}) becomes a product
 \begin{eqnarray}
R_{\rm Poisson}(k,s) &=& r(k,s)^N\label{r0}\\
r(k,s) &=& D \int dx p_b(Dx) \left(\frac{x^2-(s/2)^2}{x^2-(s/2)^2+2 i k s \lambda^2/\beta}\right)^{\beta/2} \label{r1}   \ .
\end{eqnarray}
The universal final result should be independent of the distribution of the eigenvalues of the background Hamiltonian.  The simplest choice for this distribution is 
\begin{equation}
p_b(E) \ = \ \frac{1}{\sqrt{N}} 
\left\{
\begin{array}{ll}
 1 \ , \quad  & |x| \le \sqrt{N}/2 \cr
 0 \ , \quad  & |x| > \sqrt{N}/2 \ , 
\end{array} 
\right. 
\label{e15}
\end{equation}
where $D=1/\sqrt{N}$ and $\sqrt{N}=ND$ is the length of the background spectrum. 

\begin{figure}
\begin{center}
\epsfig{figure=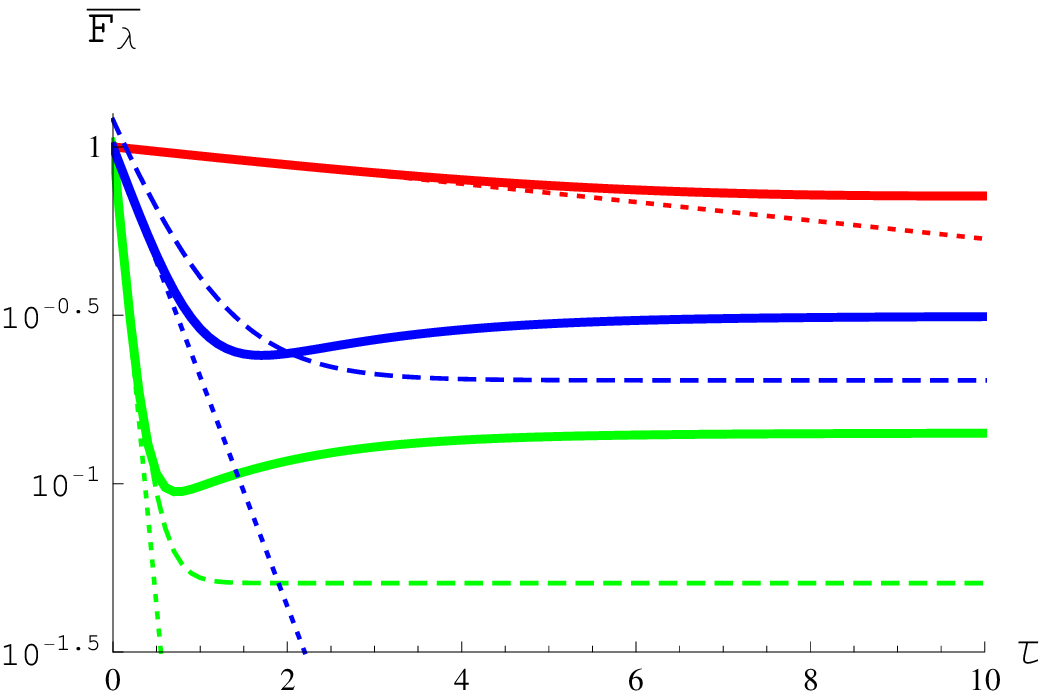, width=7cm, height=5cm}\hspace{1cm} \epsfig{figure=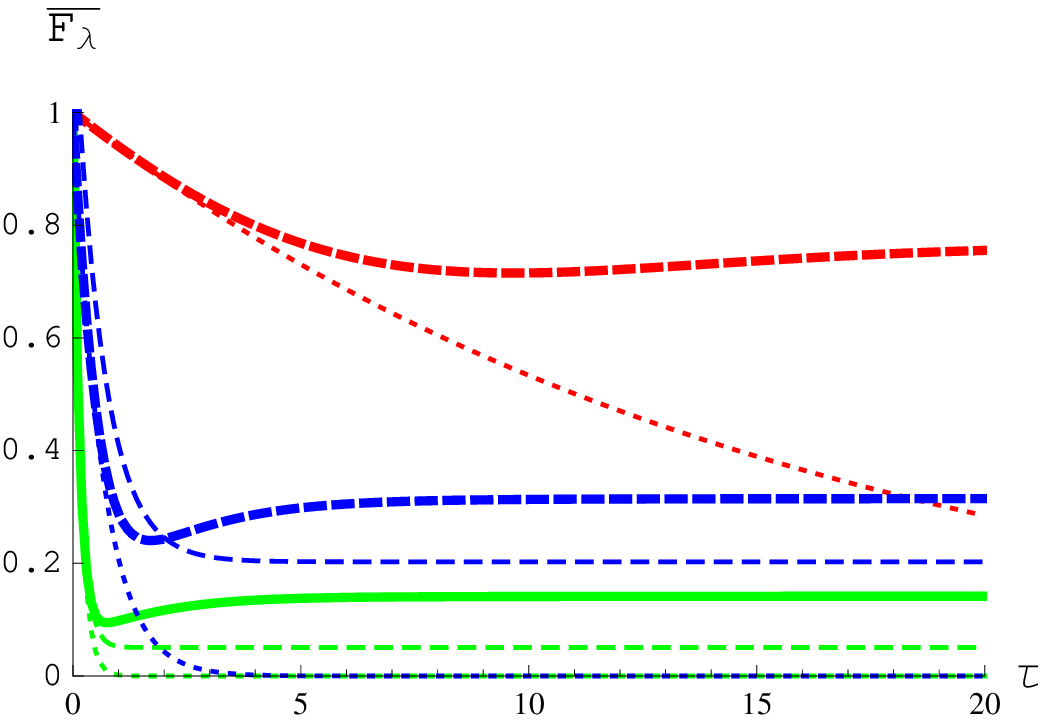, width=7cm, height=5cm}
 \caption{Plot of Eq.~(\ref{F32}) for the values $\lambda = 0.1$ (red thick line), $\lambda = 0.5$ (blue thick line) and  $\lambda = 1$
 (green thick line) on a logarithmic scale (right) and on a linear scale (left). 
The curves obtained by Fermi's golden rule are depicted by thinner dotted lines in all three cases. 
For $\lambda = 1$ and for  $\lambda = 0.5$ also the curves obtained from Eq.~(\ref{gru}) are plotted (thin dashed lines). }
\label{Poissonplot}
\end{center}
\end{figure}

For $\beta=2$ the integral (\ref{r1}) can be evaluated in the large $N$ limit by the residue theorem
\begin{eqnarray}
r(k,s)&=& 1- \frac{2\pi \lambda^2 |k||s|}{N\sqrt{s^2-4iks\lambda^2}}+{\cal O}\left(\frac{1}{N^2}\right) \ ,\\
R_{\rm Poisson}(k,s)&=& \exp\left(-\frac{2\pi \lambda^2 |k||s|}{\sqrt{s^2-4iks\lambda^2}}\right) \ .
\end{eqnarray}
Using this result together with Eq.~(\ref{r0}) and Eq.~(\ref{F0}) we find
 \begin{eqnarray}
 \label{F1}
\overline{F_\lambda(\tau)}-1 &=& {\rm Re}\int\limits_{0}^{\infty} \frac{2 ds}{\pi} 
                               \left[\cos(\tau s)-1\right] \nonumber\\
               &&\qquad\quad \int\limits_{0}^{\infty} dk \exp\left(-ik s- \frac{2\pi \lambda^2 k s}{\sqrt{s^2-4iks\lambda^2}}\right) \ ,
\end{eqnarray}
which is almost our final result. However the remaining double integral is numerically difficult due to the oscillatory terms. We proceed by rotating the contour of $k$ integration on the negative imaginary axis. Introducing the new integration variable $x= ik/s$
one arrives at
 \begin{eqnarray}
 \label{F2}
\overline{F_\lambda(\tau)} &=& 1+ \frac{1}{\pi}\int\limits_{-\infty}^{\infty} ds s
                    \left[\cos(\tau s)-1\right]\int\limits_{0}^{1/4\lambda^2} dx  
                                \exp\left(-xs^2\right)\sin\left(\frac{2\pi \lambda^2 x s}{\sqrt{1-4x\lambda^2}}\right) \ .
                             \nonumber\\ 
\end{eqnarray}
The $s$ integration can now be performed without difficulties. The final result is
 \begin{eqnarray}
 \label{F32}
\overline{F_\lambda(\tau)} &=&1+ \frac{\lambda}{2\sqrt{\pi}}\int\limits_{0}^{1} \frac{dx}{\sqrt{x}} 
                               e^{-\frac{x\pi^2\lambda^2}{4(1-x)}} 
                             \left\{\frac{\pi}{\sqrt{1-x}}\left(e^{-\frac{\tau^2\lambda^2}{x}}
                              \cosh\left(\frac{\pi\lambda^2\tau}{1-x}\right)-1\right)\right.\nonumber\\ 
                           &&  \left.\qquad\qquad-\frac{2\tau}{x}e^{-\frac{\tau^2\lambda^2}{x}}
                               \sinh\left(\frac{\pi\lambda^2\tau}{1-x}\right)\right\} \ .
\end{eqnarray}
The remaining integral does not seem to have a simple analytic solution.
As expected $\overline{\rm IPR}_\lambda$ is not zero but saturates at a finite value, given by
 \begin{eqnarray}
 \label{F3}
 \overline{{\rm IPR}_\lambda} &=& 1-
     \frac{\sqrt{\pi}^3\lambda}{2}\exp\left(\frac{\pi\lambda}{2}\right)^2{\rm erfc}\left(\frac{\pi\lambda}{2}\right) \ .
\end{eqnarray}
This function behaves for small/large values of $\lambda$ as follows
\begin{eqnarray}
\label{asyPoi}
 \overline{{\rm IPR}_\lambda} &\simeq &\left\{\begin{array}{ccc}1-\pi^{3/2}\lambda & {\rm for}& \lambda\ll 1\cr
                                           \displaystyle{\frac{2}{\pi^2\lambda^2}}& {\rm for}& \lambda\gg 1 \ .\end{array}\right.
\end{eqnarray}
In Figure \ref{Poissonplot} survival probability is plotted on a logarithmic and on a linear scale as a function of time in units of Heisenberg time for three different values of the mean coupling strength $\lambda = 0.1$, $0.5$ and $1$ corresponding to a spreading width $\Gamma/D \approx 0.06, 1.5$ and $ 6.3$. It is seen that the survival probability reaches a minimum and increases afterwards to its saturation value given in Eq.~(\ref{F3}). 
The time evolution of survival probability splits into three regimes: for $t \ll \tau_{\rm H}/\Gamma $ fidelity follows the FGR law, for $t \gg \tau_{\rm H}/\Gamma $ survival probability has approached its saturation value and is approximately constant, in the region  $t \simeq \tau_{\rm H}/\Gamma $ the time evolution is a complicated smooth function, which interpolates between the two limiting regimes. 

In Figure \ref{Poissonplot} also the curves obtained from Eq.~(\ref{gru}) are plotted. It is seen that for a Poisson distributed spectrum of the background, Eq.~(\ref{gru}) is a rather poor approximation of the exact curve. In particular it predicts no revival of survival probability and the  saturation value is underestimated by a factor four for large coupling strength $\lambda$.

\subsection{Survival probability for a chaotic  Background: GUE}
\label{sec3}
\begin{figure}
\begin{center}
\epsfig{figure=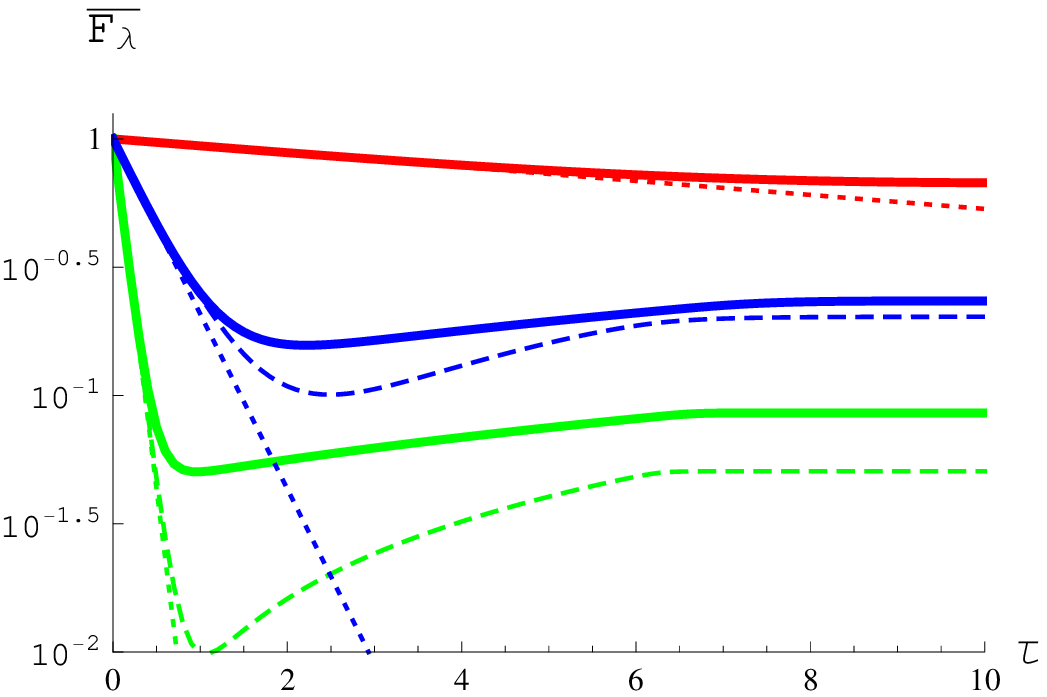, width=7cm, height=5cm}\hspace{1cm} \epsfig{figure=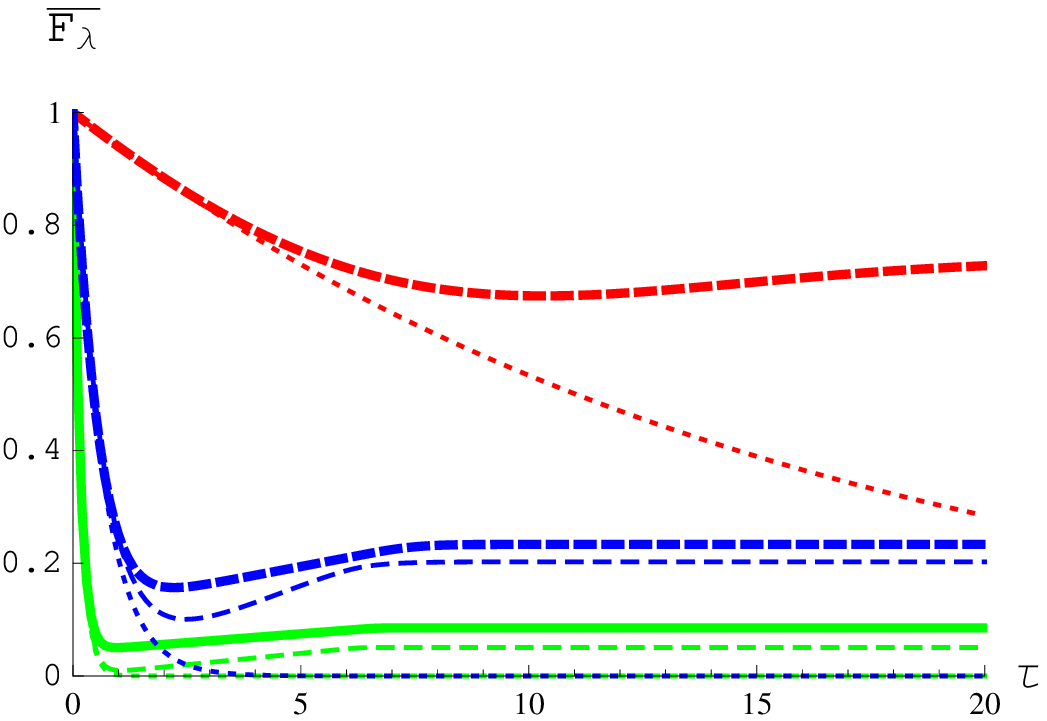, width=7cm, height=5cm}
 \caption{Plot of Eq.~(\ref{finalres}) for the values $\lambda = 0.1$ (red thick line), $\lambda = 0.5$ (blue thick line) and 
 $\lambda = 1$ (green thick line). The curves obtained by Fermi's golden rule are depicted by thinner dashed lines in all three cases. 
For $\lambda = 1$ and for  $\lambda = 0.5$ also the curves obtained from Eq.~(\ref{gru}) are plotted (thin dashed lines).}
\label{fig0}
\end{center}
\end{figure}

Using the formulae of Theorem 1.3.2 of Ref.~\cite{bor04} \footnote{For the convenience of the reader we provide Theorem 1.3.2 of Ref.~\cite{bor04} in the form needed here in \ref{AppA}.} we find for $R(k,s)$ for a chaotic background Hamiltonian with broken time reversal invariance 
\begin{eqnarray}\label{RGUE}
R_{\rm GUE}(k,s) &=& \exp\left(-i\pi\sgn(ks)\sqrt{s^2-4iks\lambda^2}\right)\nonumber\\
       &&\qquad \left( \cos(\pi s) + i \sgn(ks)\sin(\pi s)\frac{s-2ik\lambda^2}{\sqrt{s^2-4iks\lambda^2}}\right) \ .
\end{eqnarray}
We use that $R(-k,s)\ =\ R^*(k,s)$ and rotate the contour of the $k$--integral as in the Poisson case to the negative ($s>0$) or positive ($s<0$) imaginary axis. We find for the averaged survival probability 
 \begin{eqnarray}
 \label{F2a}
&&\overline{F_\lambda(\tau)} \ =\  1+ \frac{1}{\pi}\int\limits_{-\infty}^{\infty} ds s 
                    \left[\cos(\tau s)-1\right]\int\limits_{0}^{1/4\lambda^2} dx  
                                \exp\left(-xs^2\right)\nonumber\\
                   &&\ \left\{ \sin\left(\pi s \sqrt{1-4x\lambda^2}\right)\cos(\pi s)-
          \cos\left(\pi s \sqrt{1-4x\lambda^2}\right)\sin(\pi s)\frac{1-2\lambda^2 x}{\sqrt{1-4x\lambda^2}}\right\}  \ .
                             \nonumber\\ 
\end{eqnarray} 
The $s$ integration can be performed in a tedious but straightforward way. The final result is again an integral expression
\begin{eqnarray}
\overline{F_\lambda(\tau)} &= & 1+ \frac{\lambda}{2\sqrt{\pi}}\int_0^1\frac{dx}{\sqrt{x}^3\sqrt{1-x}}\exp\left(-\frac{\pi^2\lambda^2W_+^2}{x}\right)
\left(\frac{x}{2}-W_-\right)\nonumber\\ 
&&\quad\left\{ \pi W_+ \left[1-\exp\left(-\frac{\lambda^2\tau^2}{x}\right)
\cosh\left(\frac{2\pi\lambda^2\tau W_+}{x}\right)\right]\right.\nonumber\\
&&\quad\left. +\tau \exp\left(-\frac{\lambda^2\tau^2}{x}\right)\sinh\left(\frac{2\pi\lambda^2\tau W_+}{x}\right)\right\}\nonumber\\
&&\quad-\frac{\lambda}{2\sqrt{\pi}}\int_0^1\frac{dx}{\sqrt{x}^3\sqrt{1-x}}\exp\left(-\frac{\pi^2\lambda^2W_-^2}{x}\right)
\left(W_+ - \frac{x}{2}\right)\nonumber\\ 
&&\quad\left\{ \pi W_- \left[1-\exp\left(-\frac{\lambda^2\tau^2}{x}\right)
\cosh\left(\frac{2\pi\lambda^2\tau W_-}{x}\right)\right]\right.\nonumber\\
&&\quad\left. +\tau \exp\left(-\frac{\lambda^2\tau^2}{x}\right)\sinh\left(\frac{2\pi\lambda^2\tau W_-}{x}\right)\right\}\ ,
\label{finalres}
\end{eqnarray}
where
\begin{equation}
W_{\pm} \ =\ 1\pm \sqrt{1-x} \ .
\end{equation}
In Figure \ref{fig0} survival probability is plotted on a logarithmic and on a linear scale as a function of time in units of Heisenberg time for three different values of the mean coupling strength $\lambda = 0.1$, $0.5$ and $1$. These values of $\lambda$ correspond to a spreading width $\Gamma/D \approx 0.06, 1.5$ and $ 6.3$. It is seen that qualitatively the curves are quite similar to the ones obtained for a regular environment. For a GUE background energy correlations are present and a revival of survival probability is predicted by the estimation (\ref{gru}). However we notice that a revival occurs also for small values of coupling strength like $\lambda=0.1$.

The averaged inverse participation ratio is easily obtained from Eq.~(\ref{finalres}) by taking the limit  $\tau\to\infty$
\begin{eqnarray}
\label{satGUE}
\overline{{\rm IPR}_\lambda} &=& 1-\frac{\lambda\sqrt{\pi}}{2}\int\limits_0^1\frac{dx}{\sqrt{x}\sqrt{1-x}}
                                    \exp\left(-\frac{\pi^2\lambda^2(2-x)}{x}\right)\nonumber\\
               && \left[\cosh\left(\frac{2\pi^2\lambda^2\sqrt{1-x}}{x}\right)+
                \sqrt{1-x}\sinh\left(\frac{2\pi^2\lambda^2\sqrt{1-x}}{x}\right)\right] \ .
\end{eqnarray}
The remaining integral can be simplified and expressed in terms of complementary error functions akin to Eq.(\ref{F3})
\begin{eqnarray}
\label{IPRGUE}
\overline{{\rm IPR}_\lambda} & = & 
          1-\frac{2\pi^4 \lambda^4}{\sqrt{\pi}} \int\limits_0^{\infty}\frac{e^{-u^2}du}{(u^2+\pi^2 \lambda^2)^2}\nonumber\\
& = & 1 - \pi^2\lambda^2 - \frac{\pi^2 \lambda}{2\sqrt{\pi}}\left(1- 2\pi^2\lambda^2\right)
        \exp\left(\pi^2\lambda^2\right) {\rm erfc}(\pi\lambda) \ .
\end{eqnarray} 
In the limits of small and large $\lambda$ we obtain
\begin{eqnarray}
\label{asyGUE}
 \overline{{\rm IPR}_\lambda} &\simeq &\left\{\begin{array}{ccc}1-\pi^{3/2}\lambda & {\rm for}& \lambda\ll 1\cr
                                           \displaystyle{\frac{1}{\pi^2\lambda^2}}& {\rm for}& \lambda\gg 1 \ .\end{array}\right.
\end{eqnarray}
The asymptotic value of  $\overline{{\rm IPR}_\lambda}$ is half the value obtained for a regular environment but twice the value predicted by the Boltzmann--Drude approximation.

\subsection{Survival Probability for a chaotic  Background: GOE}
\label{sec4}
\begin{figure}
\begin{center}
\epsfig{figure=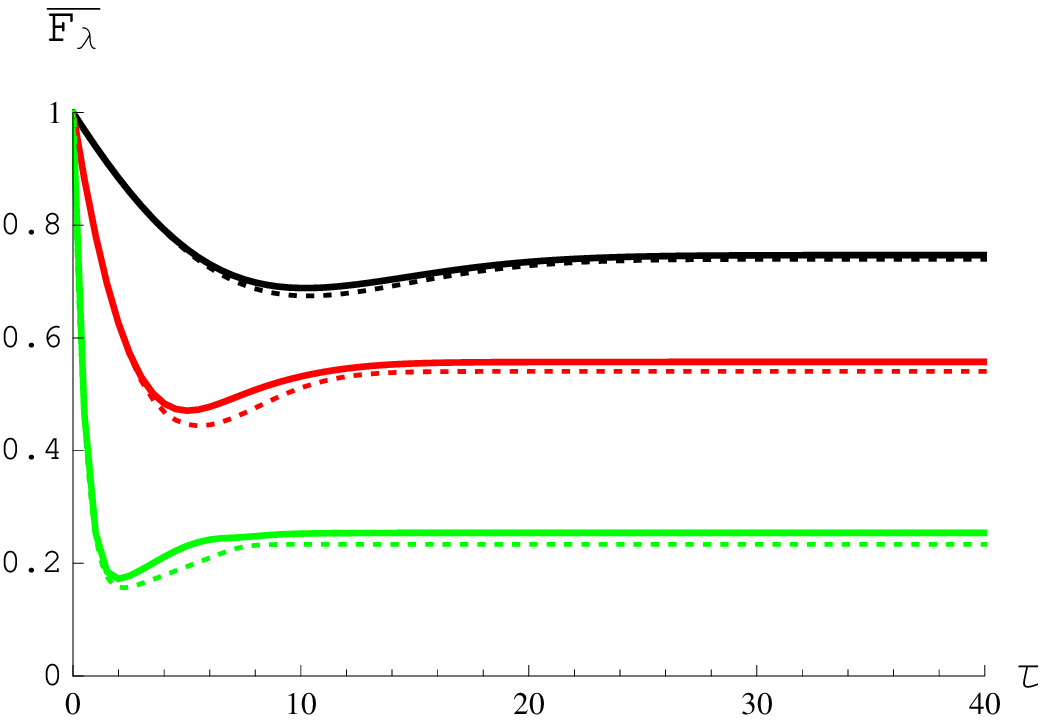, width=7cm, height=5cm}\hspace{1cm} \epsfig{figure=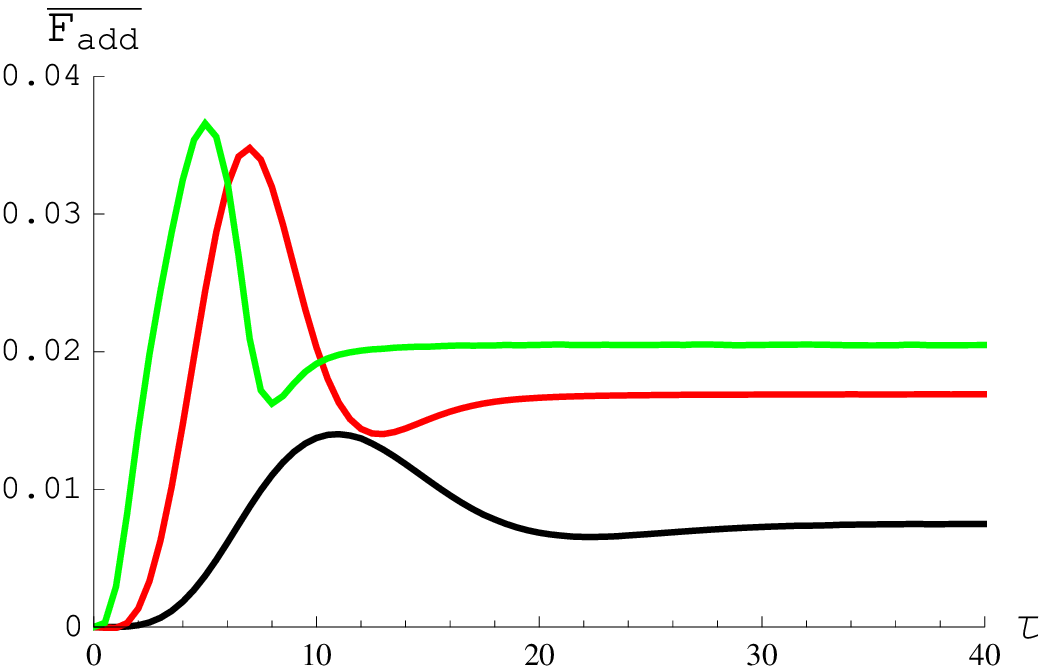, width=7cm, height=5cm}
\caption{Left: Plot of Eq.~(\ref{GOEfinal1}) and Eq.~(\ref{GOEfinal}) for the values $\lambda = 0.1$ (full black line), $\lambda = 0.2$ ( full  red line) and 
 $\lambda = 0.5$ ( full green line). For comparison the curves for the GUE are plotted for the same values of $\lambda$ 
 with dotted lines. Right: Plot of the additional ``Cooperon'' contribution to survival probability according to 
Eq.~(\ref{GOEfinal}) for the three values $\lambda=0.1$ (black line), $\lambda=0.2$ (red line) and $\lambda=0.5$ (green line).}
\label{figGOE}
\end{center}
\end{figure}

 Using Theorem 1.3.1 of Ref.\cite{bor04}, for details see \ref{AppA}, an expression for $R_{\rm GOE}(k,s)$ can be derived. We find that 
 \begin{eqnarray}
 R_{\rm GOE}(k,s) &=&  R_{\rm GUE}(k,s) + R_{\rm add}(k,s)\nonumber\\
 R_{\rm add}(k,s) &=& \frac{-i4\sgn(ks)k^2s\lambda^4}{\sqrt{s^2-4iks\lambda^2}}\left(\frac{d}{ds}\frac{\sin(\pi s)}{s}\right)
                      \times\nonumber\\
                      &&\quad\times \int\limits_1^\infty 
                      \exp\left(-i\pi \sgn(ks)\sqrt{s^2-4iks\lambda^2}t\right) \frac{dt}{t}  \ . \label{RGOE}
 \end{eqnarray}
 In the same fashion survival probability splits into two parts
 \begin{eqnarray}
\label{GOEfinal1}
 \overline{F_{\lambda, {\rm GOE}}}&=&  \overline{F_{\lambda, {\rm GUE}}}+  \overline{F_{\lambda, {\rm add}}}\nonumber\\
  \overline{F_{\lambda, {\rm add}}}&=& \int ds \int \frac{dk}{2\pi} 
                              \left[\exp\left(i\tau s)\right)-1\right]\exp\left(-i k s\right)  R_{\rm add}(k,s)\ .
 \end{eqnarray} 
The additional contribution to the survival probability $\overline{F_{\lambda, {\rm add}}}$ plays the role of a Cooperon contribution.
In a tedious but straightforward calculation we find for  $\overline{F_{\lambda, {\rm add}}}$ an expression as a double integral
\begin{eqnarray}
\overline{F_{\lambda, {\rm add}}}&=& \int\limits_0^1dx\int\limits_1^\infty dt \frac{{\sqrt{\pi x}}\lambda }{8\,t\,{\sqrt{1 - x}}}
                                     \left(H(\tau)+H(-\tau)-2H(0)\right)\nonumber\\                                     
 H(\tau)  &=& e^{-\frac{\lambda^2}{x}\left( \pi^2 + W(\tau)^2\right)}
                                   \left[\left(\frac{4\lambda^2\pi}{x}+\frac{1}{\pi}\right) W(\tau) 
                                     \sinh\left(\frac{2\,\pi \,{\lambda }^2 W(\tau)}{x}\right)\right.\nonumber\\ 
                                  && \quad\left. - \frac{2 \lambda^2}{x}\left(\pi^2 + W(\tau)^2 \right) 
                                      \cosh\left(\frac{2\,\pi \,{\lambda }^2 W(\tau)}{x}\right)\right]\nonumber\\
 W(\tau) &=& \tau+\pi t\sqrt{1-x}      \ , \label{GOEfinal}                        
 \end{eqnarray} 
which can be evaluated numerically without problems. Likewise the average of the IPR obtains an additional contribution 
\begin{eqnarray}
\overline{{\rm IPR}_{\lambda, {\rm GOE}}}&=&  \overline{{\rm IPR}_{\lambda, {\rm GUE}}}+  
                                              \overline{{\rm IPR}_{\lambda, {\rm add}}}\nonumber\\
 \overline{{\rm IPR}_{\lambda, {\rm add}}}&=& -\int\limits_0^1dx\int\limits_1^\infty dt \frac{{\sqrt{\pi x}}\lambda H(0)}
                    {4\,t\,{\sqrt{1 - x}}} \ .
\end{eqnarray}
On the left hand side of Fig.~\ref{figGOE} survival probability as given by Eq.~(\ref{GOEfinal1}) is plotted for different coupling strength 
$\lambda =$ $0.1$, $0.2$ and $0.5$ (full lines). A comparison with the corresponding curves for a GUE background (dotted lines) shows that the difference is minimal. Whether or not the background dynamics is time reversal invariant or not has no influence of the decay of the special state. Nevertheless, it is interesting to look at the Cooperon contribution   $\overline{F_{\lambda, {\rm add}}}$ separately. It is plotted on the right hand side of Fig.~\ref{figGOE} for the same values as before. We see that the contribution is small compared to $\overline{F_{\lambda, {\rm GUE}}}$. Surprisingly, we see that it is oscillating a few times with a frequency $\propto 1/\lambda$ before reaching its saturation value $\overline{{\rm IPR}_{\lambda, {\rm add}}}$ which is a non--monotonous function of $\lambda$. This contribution to the total mean IPR is depicted in the inset of Fig.~\ref{fig1} as a function of the coupling strength $\lambda$.

\subsection{Comparison}
We were able to calculate time evolution of survival probability for a complex coupling of the prepared state to a Poisson, GUE or GOE environment. As explained before a similar calculation is by now not possible for real coupling coefficents. In the latter case we resort to numerics. 

\begin{figure}
\begin{center}\epsfig{figure=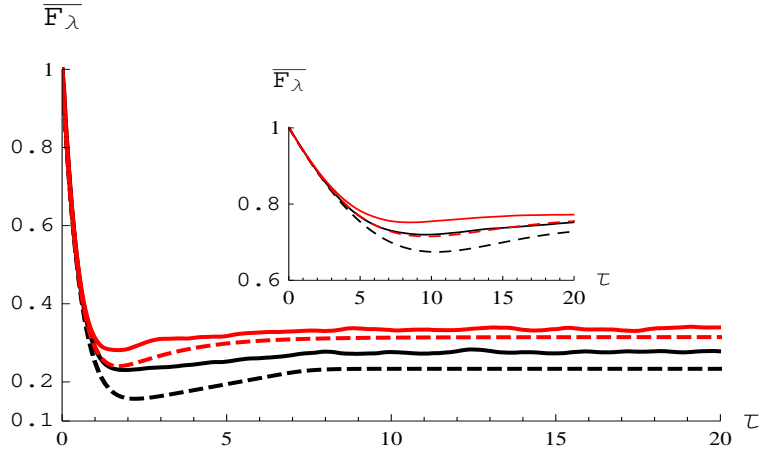, width=10cm, height=6cm}\hspace{1cm} 
 \caption{Left: Comparison of survival probability for real coupling to a Poissonian background (full red line), to a 
GUE background (full black line), for complex coupling to  a Poissonian background (dashed red line) and to a 
GUE background (dashed black line). The coupling strength is $\lambda=0.5$. The inset shows the same quantities for coupling strength $\lambda=0.1$.
\label{fig1}}
\end{center}
\end{figure}

In Fig.~\ref{fig1} time evolution of survival probability is plotted for regular and GUE background dynamics for real  and for complex coupling coefficients (the difference between a GOE background and a GUE background is almost invisible on the scale used for the plots). We see that increased background complexity reduces overall survival probability. This is in agreement with standard perturbative arguments \cite{gor06,jaq08}. 
The difference between real and complex coupling coefficients is of the same order of magnitude as the difference between regular 
and chaotic background dynamics. This is surprising inasmuch time reversal symmetry breaking in the background has practically no influence on survival probability. 

\begin{figure}
\begin{center}\epsfig{figure=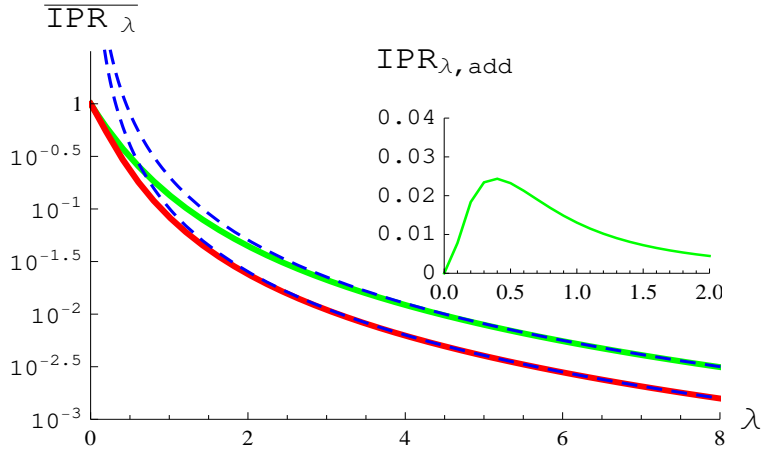, width=10cm, height=6cm}
 \caption{Left: Plot of the average inverse participation ratio $\overline{{\rm IPR}_\lambda}$ as a function of the dimensionless mean
 coupling strength $\lambda$ for complex interaction with a regular background (green curve), Eq.~(\ref{F3}), and 
for complex interaction with a GUE background (red curve), Eq.~(\ref{IPRGUE}). The blue dashed lines show the asymptotic behavior for large $\lambda$. The inset shows the additional contribution for a GOE background  $\overline{{\rm IPR}_{\lambda, {\rm add}}}$ as a function of $\lambda$.\label{fig11a}}
\end{center}
\end{figure}

In Fig.\ref{fig11a} the average inverse participation ratio is plotted for a complex coupling to a Poissonian background (as given by Eq.~(\ref{F3})) and to a GUE background (as given by Eq.~(\ref{IPRGUE}) as a function of coupling strength $\lambda$. We see that for $\lambda\gtrsim 1$ the mean IPR is well approximated by its asymptotic form, Eq.~(\ref{asyPoi}) respectively Eq.~(\ref{asyGUE}). In the inset the  additional contribution for a GOE background $\overline{{\rm IPR}_{\lambda, {\rm add}}}$ is plotted. Although negligible for practical purposes, it is interesting to see that this contribution is a non--monotonous function of $\lambda$. It vanishes for $\lambda=0$ and for  $\lambda=\infty$. It obtains its maximum for $\lambda\approx 0.5$.

\section{Discussion and Summary}
\label{dis}
We calculated exactly survival probability and fidelity amplitude for a special state, which is weakly coupled to a random matrix environment. Whereas fidelity amplitude decays exponentially according to Fermi's golden rule, survival probability shows a rich behavior.  We found a revival of survival probability after a characteristic time which increases with decreasing coupling strength and a saturation of survival probability at a value given by the mean IPR of the special state in the basis of the interacting system. Our exact results largely improve existing estimates \cite{gru97} of these quantities. 

We were able to derive analytically the full $\lambda$--dependence of the IPR in the small coupling regime, where the approximation $\overline{{\rm IPR}}\propto \lambda^{-2}$ becomes bad. It turned out that even in the strong coupling limit $\overline{{\rm IPR}}$ is largely underestimated by the Drude--Boltzmann approximation (by a factor four for a regular environment and by a factor two for a chaotic environment). 

Revival of survival probability was found for all types of background complexity. The fact that it occurs also for a regular background encumbers an explanation of the revival by spectral correlations of the background energy--levels as put forward in \cite{gru97}. The revival is quite different in nature to the fidelity revivals reported in \cite{sto04,sto05,sto06}. There a global perturbation and fidelity of a random state was considered and a revival of fidelity at Heisenberg time was found. An explanation of this phenomenon was given by the rigidity of the spectrum of a background with chaotic dynamics. Such an explanation obviously fails in the present case, since  revival occurs even in the absence of energy correlations of the background. The behavior rather resembles the overdamped oscillations in a two--level system, which is coupled to a non--Markovian heat--bath (for instance one or more spin baths \cite{pro00,bha08}) . A possible explanation is that for small couplings the system effectively reduces to a two--level system involving only the Doorway state and its nearest neighbor (in energy). This two--level system itself is then strongly coupled to the remaining background states.

Survival probability is susceptible against changes in the background dynamics from regularity to chaoticity, this is in agreement with the original arguments of Peres \cite{per84}. It is not sensitive against time reversal symmetry breaking in a chaotic background dynamics.     

Whereas fidelity amplitude has been calculated exactly in various random matrix models, this is the first exact calculation of  Fidelity in a random matrix setting. This was possible, due to advances in the calculation of ensemble averages of 
characteristic polynomials in the last years \cite{bre00,bre00a,bre01,str03,bai03,fyo03,bor04,dei00}. The exact results are limited to the case of a complex (time reversal invariance breaking) coupling to the background. A similar calculation for real (time reversal invariant) coupling would require knowledge of the averages of non--rational functions of characteristic polynomials.

\ack We thank T. Guhr, B. Gutkin, P. Mello and R. Oberhage for useful discussions. 
Two of us (HK and HJS) acknowledge support from Deutsche Forschungsgemeinschaft (DFG) within
Sonderforschungsbereich Transregio 12 ``Symmetries and Universality in
Mesoscopic Systems". HK is grateful for financial support from DFG, with  
grant No. Ko 3538/1-1 and 3538/1-2. SA acknowledges the hospitality of the University Duisburg--Essen.

\appendix
\section{Theorems 1.3.1 and 1.3.2 of Ref.~\cite{bor04} }
\label{AppA}

For a chaotic background the ensemble average can be done using a result by Borodin and Strahov \cite{bor04}. We restate Theorem 1.3.2/1.3.1 of Ref.~\cite{bor04} concerning the GUE/GOE ensemble average of a ratio of an arbitrary number of characteristic polynomials in a form adapted to our purposes.

We first state Theorem 1.3.2 concerning the GUE: Define the multivariate function ${\bf C}(\alpha,\beta)$ of $2n+2m$ variables $\alpha^-_k$, $\alpha^+_k$,  $\beta^-_l$, $\beta^+_l$, $1\leq k\leq n$, $1\leq l \leq m$ with $\alpha_k^-,\beta_l^- \in {\mathbb C}$  and $\alpha_k^+,\beta_k^+ \in {\mathbb C} \backslash {\mathbb R}$ as the ratio of an arbitrary number of characteristic polynomials
\begin{equation}
{\bf C}(\alpha,\beta)\ =\ \frac{\prod_{k=1}^n\det(H-\alpha^-_k) \prod_{l=1}^m\det(H-\beta_l^-)}{\prod_{k=1}^n\det(H-\alpha^+_k) \prod_{l=1}^m\det(H-\beta_l^+)}\ ,
\end{equation} 
and the average over $N\times N$ random matrices chosen from GUE
\begin{equation}
\overline{(\ldots)} = \int d[H] P_b  \left(\ldots\right) \ ,
\end{equation} 
where the distribution $P_b$ is given by Eq.~(\ref{e20}), with $\beta =2 $. Moreover, define $\gamma= (n+m)^2+(n-m)$ and the Vandermonde Determinant $\Delta_n(\alpha)$ $=$ $\prod_{i,j}(\alpha_i-\alpha_j)$. Then the following identity holds  
\begin{eqnarray}
\label{theounit}
\lim_{N\to \infty}{\bf C}(\alpha/\sqrt{2N},\beta/\sqrt{2N}) &=& (-1)^{\gamma/2}
\frac{\prod_{k,l}^n(\alpha_k^--\alpha_l^+)\prod_{k,l}^m(\beta_k^--\beta_l^+)}{\Delta_n(\alpha^-)\Delta_n(\alpha^+)\Delta_m(\beta^-)
\Delta_m(\beta^+)}\nonumber\\
     &&\qquad \qquad \det[S^{(2)}(\alpha^-,\beta^+|\beta^-,\alpha^+)] \ ,
\end{eqnarray}
where $S^{(2)}(\alpha^-,\beta^+|\beta^-,\alpha^+)$ is a $n+m$ matrix with rows parametrized by elements $\alpha^-_k$, $\beta^+_l$ and columns parametrized by elements
$\beta^-_k$, $\alpha^+_l$ and with matrix elements \footnote{In the second line of Eq~(\ref{FFblock}) there is a minus sign changed as compared to the original theorem of Ref.~\cite{bor04}, which is apparently wrong.} 
\begin{eqnarray}
 S^{(2)}(\alpha_p^-,\beta_q^-)&=&\frac 1\pi\, \frac{\sin(\alpha_p^--\beta_q^-)}{\alpha_p^--\beta_q^-}\,,\\
  S^{(2)}(\alpha_p^-,\alpha_q^+)&=&\left\{\begin{array}{cc}
                              -\frac{\exp{i(\alpha^+_q-\alpha_p^-)}}{\alpha_q^+-\alpha_p^-},&{\rm Im} \alpha_q^+>0,\cr
                                \frac{\exp{i(\alpha_p^--\alpha_q^+)}}{\alpha_p^--\alpha_q^+},&{\rm Im}\alpha_q^+<0,\end{array}\right.\\
S^{(2)}(\beta_p^+,\beta_q^-)&=&\left\{\begin{array}{cc}
\frac{\exp{i(\beta^+_p-\beta_q^-)}}{\beta_p^+-\beta_q^-},&{\rm Im} \beta_p^+>0,\cr
-\frac{\exp{i(\beta_q^--\beta_p^+)}}{\beta_q^--\beta_p^+},
&{\rm Im}\beta_p^+<0,\end{array}\right.\\
S^{(2)}(\beta_p^+,\alpha_q^+)&=& 2\pi i \left\{\begin{array}{cc}\frac{\exp i(\beta_p^+-\alpha_q^+)}
                                               {\beta_p^+-\alpha_q^+},&{\rm Im}\beta_p^+>0,\,{\rm Im}\alpha_q^+<0,\\
\frac{\exp i(\alpha_q^+-\beta_p^+)}{\alpha_q^+-\beta_p^+},&{\rm Im}\beta_p^+<0,\,{\rm Im}\alpha_q^+>0,\\
 0,&{\rm in\ all\ other\ cases}.\end{array}\right.\label{FFblock}
\end{eqnarray}
Since the mean level spacing is given by
 $D= \pi/\sqrt{2N}$, we find that $R(k,s)$ is exactly of the form (\ref{theounit}) with $n=m=1$ and with
 \begin{eqnarray}
 &&\alpha^-_1\ =\ \frac{\pi s}{2}\ , \qquad  \beta^-_1\ =\ - \frac{\pi s}{2}\ , \nonumber\\
 &&\alpha^+_1\ =\ \frac{\pi}{2}\sqrt{s^2-4iks\lambda^2}\ , \qquad \beta^+_1\ =\ - \frac{\pi}{2}\sqrt{s^2-4iks\lambda^2}\ .
 \end{eqnarray}
This yields Eq.~(\ref{RGUE}).

We now turn to Theorem 1.3.1 concerning the GOE: Define the multivariate function ${\bf C}(\alpha,\beta)$ of $n+m$ variables $\alpha_k$, $\beta_l$, $1\leq k\leq n$, $1\leq l\leq m$ with $\alpha_k\in {\mathbb C}$  and $\beta_l \in {\mathbb C} \backslash {\mathbb R}$ as the ratio of an arbitrary number of characteristic polynomials
\begin{equation}
{\bf C}(\alpha,\beta)\ =\ \frac{\prod_{k=1}^n\det(H-\alpha_k)}{\prod_{l=1}^m\det(H-\beta_l)}\ ,
\end{equation} 
and the average over $2N\times 2N$ random matrices chosen from GOE
\begin{equation}
\overline{(\ldots)} = \int d[H] P_b  \left(\ldots\right) \ ,
\end{equation} 
where the distribution $P_b$ is given by Eq.~(\ref{e20}), with $\beta= 1$. Then the following identity holds\footnote{a scaling factor $\sqrt{2}$ seems to be wrong in Ref.~\cite{bor04}}. 
\begin{eqnarray}
\label{theoorth}
\lim_{N\to \infty}{\bf C}(\alpha/\sqrt{4N},\beta/\sqrt{4N}) &=& 
\frac{\prod_{k=1}^n\prod_{l=1}^m(\alpha_k-\beta_l)}{\Delta_n(\alpha)\Delta_m(\beta)}\nonumber\\
     &&\qquad \qquad {\rm Pf}[S^{(1)}(\alpha,\beta|\alpha,\beta)] \ ,
\end{eqnarray}
where $S^{(1)}(\alpha^-,\beta^+|\beta^-,\alpha^+)$ is a skew--symmetric $n+m$ matrix with rows and columns parametrized by elements $\alpha$, $\beta$ and with matrix elements 
\begin{eqnarray}
S^{(1)}(\alpha_p,\alpha_q) &=&-\frac{1}{\pi}\frac{\partial}{\partial \alpha_i}\frac{\sin(\alpha_p-\alpha_q)}{\alpha_p-\alpha_q}\,,\\
S^{(1)}(\alpha_p,\beta_q) &=&\left\{\begin{array}{cc} 
-\frac{\exp{i(\beta_q-\alpha_p)}}{\beta_q-\alpha_p},&{\rm Im} \beta_q>0,\\
\frac{\exp{i(\alpha_p-\beta_q)}}{\alpha_p-\beta_q},
 &{\rm Im}\beta_q<0,
\end{array}\right.
\\
S^{(1)}(\beta_p,\beta_q)&=&2\pi i\left\{\begin{array}{cc} 
\int_{1}^{+\infty}\frac{\exp(i(\beta_p-\beta_q)t)}t\,dt,&{\rm Im}
\beta_p>0,\,{\rm Im}\beta_q<0,\\
-\int_{1}^{+\infty}\frac{\exp(i(\beta_q-\beta_p)t)}t\,dt,&{\rm Im}
\beta_p<0,\,{\rm Im}\beta_q>0,\\
0,&{\rm in\ all\ other\ cases}.\end{array}\right.
\end{eqnarray}
Setting 
 \begin{eqnarray}
 &&\alpha_1\ =\ \frac{\pi s}{2}\ , \qquad  \alpha_2\ =\ - \frac{\pi s}{2}\ , \nonumber\\
 &&\beta_1\ =\ \frac{\pi}{2}\sqrt{s^2-4iks\lambda^2}\ , \qquad \beta_2\ =\ - \frac{\pi}{2}\sqrt{s^2-4iks\lambda^2}\ .
 \end{eqnarray}
in Eq.~(\ref{theoorth}) yields Eq.~(\ref{RGOE}).
\section*{References}
\providecommand{\newblock}{}


\begin{thebibliography}{10}
\expandafter\ifx\csname url\endcsname\relax
  \def\url#1{{\tt #1}}\fi
\expandafter\ifx\csname urlprefix\endcsname\relax\def\urlprefix{URL }\fi
\providecommand{\eprint}[2][]{\url{#2}}

\bibitem{nie00}
Nielsen M~A and Chuang I~L 2000 {\em Quantum Computation and Quantum
  Information\/} (Cambridge: Cambridge University Press)

\bibitem{gor06}
Gorin T, Prosen T, Seligman T~H and Znidaric M 2006 {\em Phys. Rep.\/} {\bf
  435} 33

\bibitem{jaq08}
Jaquod P and Petitjean C 2009 {\em Adv. Phys.\/} {\bf 58} 67

\bibitem{bog06}
Bogomolny E and {\em et al} 2006 {\em Phys. Rev. Lett.\/} {\bf 97} 254102

\bibitem{akk06}
Akkermans E and Montambaux G 2006 {\em Mesoscopic Physics of electrons and
  photons\/} 1st ed (Cambridge: University Press)

\bibitem{pri94}
Prigodin V~N, Altshuler B~L, Efetov K~B and Iida S 1994 {\em Phys. Rev.
  Lett.\/} {\bf 72} 546

\bibitem{gut09}
Guti{\'e}rrez M, Waltner D, Kuipers J and Richter K 2009 {\em Phys. Rev. E\/}
  {\bf 79} 046212

\bibitem{sie01}
Sieber M and Richter K 2001 {\em Physica Scripta\/} {\bf T90} 128

\bibitem{mue04}
M{\"u}ller S, Heusler S, Braun P, Haake F and Altland A 2004 {\em Phys. Rev.
  Lett.\/} {\bf 93} 014103

\bibitem{gru97}
Gruver J~L, Aliagla J, Cerdeira H~A, Mello P~A and Proto A~N 1997 {\em Phys.
  Rev. E\/} {\bf 55} 6370

\bibitem{flo98}
Flores J, Hernandez-Saldana H, Leyvraz F and Seligman T~H 1998 {\em J. Phys.
  A\/} {\bf 31} 1509

\bibitem{wei30}
Weisskopf V and Wigner E~P 1930 {\em Z. Phys.\/} {\bf 63} 54

\bibitem{meh04}
Mehta M~L 2004 {\em Random Matrices\/} 3rd ed (Amsterdam: Elsevier)

\bibitem{bor04}
Borodin A and Strahov E 2004 {\em {Communication on Pure and Applied
  Mathematics}\/} {\bf 37} 209

\bibitem{boh69}
Bohr A and Mottelson B 1969 {\em Nuclear Structure Vol. 1\/} 9th ed (New York:
  W. A. Benjamin)

\bibitem{guh98}
Guhr T, M{\"u}ller-Gr{\"o}ling A and Weidenm{\"u}ller H~A 1998 {\em Phys.
  Rep\/} {\bf 299} 189

\bibitem{stoe99}
St{\"o}ckmann H~J 1999 {\em Quantum Chaos an Introduction\/} 1st ed (Cambridge:
  University Press)

\bibitem{mel95}
Mello P~A 1995 {\em Mesoscopic Quantum Physics\/} (Elsevier)

\bibitem{koh09}
Kohler H, Guhr T and {\AA}berg S 2009 To be published

\bibitem{bre00}
Brezin E and Hikami S 2000 {\em Comm. Math. Phys.\/} {\bf 214} 111

\bibitem{bre00a}
Brezin E and Hikami S 2000 {\em Phys. Rev. E\/} {\bf 62} 3558

\bibitem{bre01}
Brezin E and Hikami S 2001 {\em Comm. Math. Phys.\/} {\bf 223} 363

\bibitem{str03}
Strahov E and Fyodorov Y 2003 {\em Comm. Math. Phys.\/} {\bf 241} 343

\bibitem{bai03}
Baik Y, Deift P and Strahov E 2003 {\em J. Math. Phys.\/} {\bf 44} 3657

\bibitem{fyo03}
Fyodorov Y~V and Keating J~P 2003 {\em J. Phys. A\/} {\bf 36} 4035

\bibitem{dei00}
Deift P 2000 {\em Orthogonal Polynomials and Random Matrices: A
  Riemann--Hilbert Approach\/} Courant lecture notes 3 (New York: New York
  University)

\bibitem{sto04}
St{\"o}ckmann H~J and Sch{\"a}fer R 2004 {\em New J. Phys.\/} {\bf 6} 199

\bibitem{sto05}
St{\"o}ckmann H~J and Sch{\"a}fer R 2005 {\em Phys. Rev. Lett.\/} {\bf 94}
  244101

\bibitem{sto06}
St{\"o}ckmann H~J and Kohler H 2006 {\em Phys. Rev. E\/} {\bf 73} 066212

\bibitem{pro00}
Prokof'ev N~V and Stamp P~C~E 2000 {\em Rep. Prog. Phys.\/} {\bf 63} 669

\bibitem{bha08}
Rao D~D~B, Kohler H and Sols F 2008 {\em New Jour. Phys.\/} {\bf 10} 115017

\bibitem{per84}
Peres A 1984 {\em Phys. Rev. A\/} {\bf 30} 1610

\end{thebibliography}
\end{document}